\documentclass[prd,superscriptaddress,twocolumn]{revtex4}

\usepackage{caption}
\usepackage{subcaption}
\usepackage{amssymb}
\usepackage{amsfonts}
\usepackage{amsmath}
\usepackage{graphicx}
\usepackage{dcolumn}
\usepackage{bm}
\usepackage[colorlinks]{hyperref}
\usepackage{soul} % Para usar resaltado

\newif\ifhighlight
\highlightfalse % Cambiar esto a \highlightfalse para desactivar el resaltado

\newcommand{\correction}[1]{\ifhighlight\hl{#1}\else#1\fi}

\begin{document}
\title{\correction{Exact results in a punctured neighborhood of a strong curvature singularity}}

\author{Franco Fiorini}
\email{francof@cab.cnea.gov.ar}
\affiliation{Grupo de Comunicaciones Ópticas, Departamento de Ingeniería en Telecomunicaciones, Consejo Nacional de Investigaciones Científicas y Técnicas (CONICET) and Instituto Balseiro (UNCUYO), Centro Atómico Bariloche, Av.~Ezequiel Bustillo 9500, CP8400, S. C. de Bariloche, Río Negro, Argentina}

\author{Juan Manuel Paez}
\email{juan.paez@ib.edu.ar}
\affiliation{Grupo de Comunicaciones Ópticas, Departamento de Ingeniería en Telecomunicaciones, Consejo Nacional de Investigaciones Científicas y Técnicas (CONICET) and Instituto Balseiro (UNCUYO), Centro Atómico Bariloche, Av.~Ezequiel Bustillo 9500, CP8400, S. C. de Bariloche, Río Negro, Argentina}

\begin{abstract}
By constructing a model of spacetime having a strong curvature singularity in which causal geodesics are complete, but more generic causal curves are not, we explicitly show that some electrostatic field configurations on that background are bounded on every open punctured neighborhood of the curvature singularity. Our calculations are performed using the  \correction{analog} gravity description provided by Plebanski and Tamm, according to which the characterization of the electromagnetic field on a generic curved background is equivalent to solving Maxwell's equations in flat space with a matter content verifying certain nontrivial constitutive relations. The regularity of the electric field as it approaches what could be considered the worst conceivable physical condition, opens the door to further investigation into the possibility of propagating signals capable of crossing a spacetime singularity.  
\end{abstract}
\maketitle
\section{Introduction}

Spacetime singularities are inevitable consequences of gravitational theories constructed on the basis of pseudo-Riemannian geometry. As discovered in the 1970s, their occurrence manifests not only in a large class of solutions in the context of general relativity, but virtually in any geometrically conceived framework describing gravity \cite{HP}. Plainly, a singularity is the end point of an incomplete causal curve, which in the specific case of a ``strong curvature" singularity, represents the end of the world line of a particle in an event characterized by the unbounded growth of some scalar constructed from the curvature tensor. This is what happens  in the paradigmatic well-known singularities present in the interior of most black hole solutions and in the universal singular state representing the big bang in a variety of cosmological models. In the black hole case, not every causal curve or geodesic hits the singularity, but certainly an infinite number of them actually do. For instance, in the simple example provided by the maximal analytical extension of the Schwarzschild black hole, the curvature singularity lies in the future of all causal curves inside the event horizon. Hence, every causal curve crossing the event horizon will inevitably hit the singularity. Antithetically, the cosmological singularity has a universal character because any causal geodesic is incomplete if its affine parameter (the proper time $t$ in the case of timelike geodesics) is sufficiently extended into the past. 

Many other types of singularities are nowadays studied and classified, including a considerable amount appearing in circumstances not associated with divergences of curvature scalars, as in the case of quasiregular singularities. The majority of key results in the area until the end of the past century are summarized in the standard textbooks \cite{HE}-\cite{Naber}, while later research is covered in the nice review articles \cite{Senovilla} and \cite{J2}. More recently, a number of studies dealing with the possibility of crossing singularities, and thus connecting what otherwise were causally disconnected regions of the spacetime, show that the research in the area is still quite active, see, e.g., \cite{singularitycrossing1} and \cite{singularitycrossing2}. These developments are not only important from a conceptual point of view but also crucial, for instance, to support cosmological models where the present universe is powered by pre-Big Bang physics, as seems to be the case in conformal cyclic cosmology \cite{Penrose}; see also \cite{ccc}.

Singularities are not circumscribed to the behavior of causal curves and geodesics alone; sometimes, they are not necessarily associated with divergences of physical quantities related to the presence of matter fields in the spacetime, even though this is usually the case. For instance, there are geodesically complete cosmological spaces in which the pressure of the matter fields is divergent at some event, as in some cases of \emph{sudden} singularities \cite{Barrow}, \cite{Ruth}. Early examples \cite{Shepley}, more closely aligned with our present concerns, show that some curvature singularities could leave no imprints on the matter fields acting as sources of the Einstein field equations. It is for this reason that, if our purpose is to assess the possibility that some signal or influence could traverse a curvature singularity, it is important to move away from the mere study of causal curves and geodesics. Bearing this intention in mind, our work is a tentative contribution to this subject by examining the behavior of an electromagnetic (EM) field in a background containing a strong curvature singularity.    

The study of the Maxwell field on curved backgrounds has a long history by itself. Classic, early references are \cite{Pleb} and \cite{Cohen}, which settled the language used later in the late 1970s and early 1980s in the realm of astrophysics, mostly in relation to the study of accretion-disk magnetospheres in pulsars and black holes; see, for instance, \cite{Blandford}-\cite{Thorne2}. More modern and formal developments were considered in \cite{Tsagas} and \cite{Lobo}, and it is clear that the subject is far from over, as evidenced by some recent progress alluding to the propagation of guided waves in the spacetime, see \cite{Nuevo}.

Our approach to the characterization of the EM field is slightly different. We will make use of the Plebanski-Tamm (PT) formalism, which allows us to solve Maxwell's equations in a curved background as if they were the \emph{usual} EM equations in a flat-space material medium with peculiar constitutive relations. This ``gravito-optic" correspondence is framed in the context of  \correction{analog} gravity models; see, e.g., \cite{Analog1} and \cite{Tanos}. It has at least two benefits: on the one hand, it provides a clear and physically more transparent 3D visualization of inherently 3+1-dimensional phenomena. On the other, it tends a bridge between gravitational physics and its potential experimental testing in the \correction{laboratory}. Keen discussions on the actual experimental feasibility of this approach can be found in \cite{Barcelo}.

This article is structured as follows. In Section \ref{secpt}, we briefly review the gravito-optic PT analogy and discuss a couple of facts related to the propagation of waves in such a context. The toy model representing the spacetime hosting a strong, naked curvature singularity is introduced and studied in detail in Section \ref{latres}. Section \ref{estatica} is entirely dedicated to solving the electrostatic PT equations in the material medium provided by the toy model constructed in the previous section. Our notation is as follows: vector and second rank tensor components will be denoted using Greek indices $v_{\mu}$ and $g_{\mu\nu}$, respectively, where $\mu,\nu:0,1,2,3$, or Latin indices, $v_{i}$ and $g_{ij}$, if $i,j:1,2,3$. Three-dimensional vector objects will, in turn, be denoted by $\bar{v}$, and their scalar, vector, and tensor products as $\bar{v}\cdot\bar{w}$, $\bar{v}\times \bar{w}$, and $\bar{v}\otimes\bar{w}$, respectively. Scalar products of row and column vectors will be written without dots, as in $\bar{v}^{\intercal}\bar{w}$, where $^{\intercal}$ means transposition. Moreover, 3D matrices will be written as $\textbf{K}$. Cartesian vector and matrix components will be written \correction{as} $v_{i}$ and $K_{ij}$, respectively. The product of a matrix $\textbf{K}$ by a vector $\bar{v}$ will be simply $\textbf{K}\bar{v}$. All the components in 3D will be referred to a Cartesian coordinate system $\bar{x}=(x_{1},x_{2},x_{3})$ or $\bar{x}=(x,y,z)$, as in Sections \ref{latres} and \ref{estatica}. Finally, we are using units in which $c=(\epsilon_{0} \mu_{0})^{-1/2}=1$.

\section{Plebanski-Tamm media}\label{secpt}
Classical electromagnetism without sources in a four-dimensional curved background with line element $ds^2=g_{\mu\nu}dx^{\mu}dx^{\nu}$ is described by the covariant Maxwell source-free equations
\begin{equation}
F^{\mu\nu}_{\,\,\,\,\,\,\,;\mu}=0\,,\,\,\,\,(\epsilon^{\mu\nu\rho\sigma}F_{\rho\sigma})_{,\mu}=0\,, \label{cov}
\end{equation}
where $F_{\mu\nu}=A_{\nu,\mu}- A_{\mu,\nu}$, and $A_\mu=(\phi,\bar{A})$ is the four-potential (semicolons denote covariant derivative). It is a curious fact, first discovered by Tamm \cite{Tamm} and later on, in a seemingly independent manner, by Plebanski \cite{Pleb}, that, for stationary metrics, the system in Eq. (\ref{cov}) \correction{is exactly the same as} the usual three-dimensional, source-free Maxwell equations in the presence of matter in flat space, provided appropriate constitutive equations are introduced in a noncovariant way. In fact, written in Cartesian coordinates, the equations are  
\begin{eqnarray}
    &&\bar{\nabla} \cdot \bar{D}=0,\,\,\,\,\,\,\,\bar{\nabla} \times \bar{H}-\frac{\partial \bar{D}}{\partial t}=0,\label{eq:maxwell_mediosfuentescom} \\
   &&\bar{\nabla} \cdot \bar{B}=0,\,\,\,\,\,\,\,\,\,\,\,\,\, \bar{\nabla} \times \bar{E}+\frac{\partial \bar{B}}{\partial t}=0,
     \label{eq:maxwell_medioscom}
\end{eqnarray}
where the constitutive relations linking the various fields in the material medium are
\begin{equation}
\bar{D}=\textbf{K}\bar{E}+\bar{\Gamma}\times\bar{H},\,\,\,\,\,\,
\bar{B}=\textbf{K}\bar{H}-\bar{\Gamma}\times\bar{E}\,.\label{eq_PT_vector_Bcom}
\end{equation}
The analogy is valid only if the Cartesian components of the matrix $\textbf{K}$ and the vector $\bar{\Gamma}$ are related to the spacetime geometry according to
\begin{equation}
 K_{ij}=-\frac{\sqrt{-g}}{g_{00}}g^{ij}\,,\,\,\,\,\,\,\,\,
\Gamma_i=\frac{g_{0i}}{g_{00}}\,.\label{matyvec}
\end{equation}
In this way, the propagation of electromagnetic waves in a given curved spacetime (not necessarily a solution of Einstein field equations) is equivalent to the study of waves in a 3D material medium described by the constitutive equations of Eq. (\ref{eq_PT_vector_Bcom}); we refer the interested reader to \cite{leon-phil} and the appendix of \cite{leon-phi2} for a complete treatment \correction{and a thorough derivation of Eqs.} (\ref{eq:maxwell_mediosfuentescom})-(\ref{matyvec}). \correction{We should emphasize that this equivalence is valid only in the context of stationary metrics.} Moreover, it is manifestly noncovariant because it is valid only in Cartesian coordinates fixed at the material medium which is stationary; the metric components, as well as the inverse metric components appearing in Eq. (\ref{matyvec}), are all written in Cartesian coordinates and are time independent. However, covariant extensions to general media were developed, see, e.g., \cite{PlebCOV1}-\cite{Thomson}; there it was shown, among many other things, that PT equations are equivalent to the ones coming from the more general covariant approach in those cases where the material medium is stationary.  

Before proceeding, we can comment on some limits that emerge as consequences of the full PT equations. In order to do so let us consider waves of the form
\begin{eqnarray}
\bar{E}(\bar{x},t)&=&\bar{E}_{0}(\bar{x})\exp\big[i\,k_{0}\,(\bar{k}(\bar{x})\cdot\bar{x}- t)\big],\notag\\
\bar{H}(\bar{x},t)&=&\bar{H}_{0}(\bar{x})\exp\big[i\,k_{0}\,( \bar{k}(\bar{x})\cdot\bar{x}- t)\big],\label{expEyH2}
\end{eqnarray}
where $\bar{E}_{0}(\bar{x})$ and $\bar{H}_{0}(\bar{x})$ are the complex amplitudes and $\bar{k}(\bar{x})$ is the complex wave vector, all of them depending solely on the spatial coordinates $\bar{x}$. After considering this ansatz in the curl equations given by Eqs. (\ref{eq:maxwell_mediosfuentescom}) and (\ref{eq:maxwell_medioscom}), and taking into account the constitutive relations in Eq. (\ref{eq_PT_vector_Bcom}), we get 
\begin{align}
&i\,k_{0}^{-1}\,\bar{\nabla}\times\bar{E}_{0}=(\bar{\nabla}(\bar{k}\cdot\bar{x})+\bar{\Gamma})\times\bar{E}_{0}-\textbf{K} \bar{H}_{0}\notag\,, \\
&i\,k_{0}^{-1}\,\bar{\nabla}\times\bar{H}_{0}=(\bar{\nabla}(\bar{k}\cdot\bar{x})+\bar{\Gamma})\times\bar{H}_{0}+\textbf{K} \bar{E}_{0}\label{eq:rot_H_cuasi_planas}\,.
\end{align}
If, in turn, we pay attention to the divergence equations in Eqs. (\ref{eq:maxwell_mediosfuentescom}) and (\ref{eq:maxwell_medioscom}), we obtain
\begin{align}
   &-i\,k_0\left[\bar{\nabla}(\bar{k}\cdot\bar{x})+\bar{\Gamma}\right]\cdot\textbf{K}\bar{E_0}-(\bar{\nabla}\times\bar{\Gamma})\cdot\bar{H_0}= \bar{\nabla}\cdot\textbf{K}\bar{E_0}, \notag\\ 
    &-i\,k_0\left[\bar{\nabla}(\bar{k}\cdot\bar{x})+\bar{\Gamma}\right]\cdot\textbf{K}\bar{H_0}+(\bar{\nabla}\times\bar{\Gamma})\cdot\bar{E_0}=\bar{\nabla}\cdot\textbf{K}\bar{H_0}.\label{eq:div_B_igual_0}
\end{align}
Despite their apparent complexity, Eqs. (\ref{eq:rot_H_cuasi_planas}) and (\ref{eq:div_B_igual_0}) are not independent. Actually, it is shown in Appendix \ref{ap1} that, by properly combining Eq. (\ref{eq:rot_H_cuasi_planas}) with Eq. (\ref{eq:div_B_igual_0}), we obtain an identity. Of course, this is not an intrinsic, general property of PT equations, but rather of the quasi-plane-wave ansatz specified in Eq. (\ref{expEyH2}).     

The usual circumstance of having plane waves in flat, empty space is recovered from Eqs. (\ref{eq:rot_H_cuasi_planas}) and (\ref{eq:div_B_igual_0}) by taking $\textbf{K}=\textbf{I}$ and $\bar{\Gamma}=\bar{0}$, corresponding to Minkowski spacetime with metric $g_{\mu\nu}=diag(-1,1,1,1)$. Additionally, we have to consider constant $\bar{E_0}$, $\bar{H_0}$ and $\bar{k}$. Under \correction{these} requirements, the system in Eq. (\ref{eq:rot_H_cuasi_planas}) becomes
\begin{equation}
    \bar{H}_0=\bar{k}\times\bar{E}_0\,,\,\,\,\,\, \bar{E}_0=-\bar{k}\times\bar{H}_0\,.\label{eq:analogo_mink_2}\\   
\end{equation}
Coming back to the comment made in the preceding paragraph, it is straightforward to show that Eq. (\ref{eq:div_B_igual_0}) give\correction{s} rise to the conditions $\bar{k}\cdot\bar{E}_0=0$ and $\bar{k}\cdot\bar{H}_0=0$, which are not independent of Eq. (\ref{eq:analogo_mink_2}). Thus we see, on one hand, that $\bar{k}\perp\bar{E}_0$, $\bar{k}\perp\bar{H}_0$ and $\bar{E}_0\perp\bar{H}_0$, and on the other, combining Eqs. (\ref{eq:analogo_mink_2}),   
\begin{equation}
    \bar{E}_0=-\bar{k}\times(\bar{k}\times\bar{E}_0)=k^2\bar{E}_0\,.
\end{equation}
This, of course, implies $\bar{k}\cdot\bar{k}=k^2=1$, which is no other than the usual dispersion relation. 

We can also briefly discuss under what circumstances the geometrical optics regime comes out from this formalism. Even though not directly related to our present concerns, this limit is important because it illustrates that null geodesics in the spacetime are equivalent to light rays in the material medium, a very well-know\correction{n} feature used in many works to simulate in the \correction{laboratory} the effects of light propagation occurring in various spacetimes of astrophysical interest. See, for instance, \cite{Mackay1}-\cite{Turner}. For more insight into the following succinct exposition, we refer the reader to \cite{nos}.   

The geometrical optics realm involves slow variations of the fields in scales of order $k_{0}$. Additionally, the vector $\bar{\nabla}(\bar{k}\cdot\bar{x})$ is approximated by $\bar{k}$ itself. Once we incorporate this information into Eqs. (\ref{eq:rot_H_cuasi_planas}) and (\ref{eq:div_B_igual_0}), and after defining $\bar{p}=\bar{k}+\bar{\Gamma}$, we get
\begin{align}
&\bar{p}\times\bar{E}_{0}=\textbf{K} \,\bar{H}_{0}\,,\,\,\,\,\,
\bar{p}\times\bar{H}_{0}=-\textbf{K} \,\bar{E}_{0}\label{eva10}\,,\\
&\bar{p}\cdot\textbf{K} \,\bar{E}_{0}=0\,,\,\,\,\,\,
\bar{p}\cdot\textbf{K} \,\bar{H}_{0}=0\,.\label{eva20}
\end{align}
Again, we see that the divergence equations (\ref{eva20}) are not independent. Some matrix and vector handling allows us to combine Eqs. (\ref{eva10}) to obtain \begin{equation}\label{eq:f3}
\left\lbrace\big(\bar{p} \otimes \bar{p}\big)\,\textbf{K} +\Big[ \det(\textbf{K}) -  \bar{p}^{\,\intercal} \textbf{K}\,\bar{p} \,
\Big] \textbf{I}\,\right\rbrace\bar{E}_0 = \bar{0}\,,
\end{equation}
which has nontrivial solutions if and only if
\begin{equation}\label{hamilton}
H\doteq\det(\textbf{K}) -\bar{p}^{\,\intercal} \textbf{K}\,\bar{p}=0.
\end{equation}
It turns out that the identically \correction{vanishing} function $H$ can be seen as the Hamiltonian governing light-ray propagation in the optical medium, see \cite{Hamop}, also \cite{Sluijter1} and \cite{Sluijter2}. Moreover, in Ref. \cite{nos}, it was shown that the canonical equations 
\begin{equation}
\bar{\nabla}_{\bar{x}} H =-\frac{d \bar{k}}{dt},\,\,\,\,\,
\bar{\nabla}_{\bar{k}}H =\frac{d \bar{x}}{dt}, \label{hamiltonec}
\end{equation}
are susceptible to analytical treatment, where $\bar{\nabla}_{\bar{k}}\equiv (\partial/\partial k_1,\partial/\partial k_2,\partial/\partial k_3)$. Suffice to say that Eqs. (\ref{hamiltonec}) can be written solely in terms of $\textbf{K}$ and $\bar{p}$ as
\begin{eqnarray}
\frac{d \bar{x}}{dt}&=&-2 \textbf{K}\bar{p},\label{hamiltoncoorfin}\\
\frac{d \bar{k}}{dt}&=&\bar{p}^{\,\intercal}\Big[ [\textbf{K}_{i}-tr(\textbf{K}^{-1}\textbf{K}_{i})\,\textbf{K}]\,\bar{p}+2\textbf{K}\,\bar{p}_{i}\Big]\hat{e}_{i}\,,\label{hamiltonmomfin}
\end{eqnarray}
where $\hat{e}_{i}$ are the Cartesian unit vectors in $\mathbb{R}^3$, $tr(\textbf{A})$ is the trace of the matrix $\textbf{A}$, and $\textbf{K}_{i}$ are three matrices whose components are obtained from the components of $\textbf{K}$ by differentiating with respect to the coordinate $x_{i}$. 

\bigskip

In this paper we are interested in the behavior of static fields on a background space. Let us now examine what the structure of the equations is in this case. By dropping the time-varying terms in Faraday's and Ampère-Maxwell's laws, we have that $\bar{E}$ and $\bar{H}$ can be written in terms of scalar potentials $\phi_{E}$ and $\phi_{H}$ in the usual way, i.e., $\bar{E}=-\bar{\nabla}\phi_{E}$ and $\bar{H}=-\bar{\nabla}\phi_{H}$. These can be combined with the constitutive equations given in Eq. (\ref{eq_PT_vector_Bcom}) and the two divergence equations $\bar{\nabla} \cdot \bar{D}=0$ and $\bar{\nabla} \cdot \bar{B}=0$, to get
\begin{eqnarray}
    &&\bar{\nabla} \cdot (\textbf{K}\bar{\nabla}\phi_{E})+ (\bar{\nabla} \times\bar{\Gamma})\cdot\bar{\nabla}\phi_{H}=0\, ,\notag\\ 
     &&\bar{\nabla} \cdot (\textbf{K}\bar{\nabla}\phi_{H})- (\bar{\nabla} \times\bar{\Gamma})\cdot\bar{\nabla}\phi_{E}=0\,. \label{estaticac}
\end{eqnarray}
As a consequence of the electric-magnetic coupling in the constitutive equations due to the presence of a nonconstant $\bar{\Gamma}$, static $\bar{E}$ and $\bar{H}$ fields in PT media generally interact with each other. This ultimately relies on the fact that $\bar{\Gamma}$ is associated \correction{with} the off-diagonal components of the metric, which, in turn, are normally—although not always—representative of axisymmetric spacetimes endowed with some sort of rotation. Unlike in the previous cases, where quasi-plane-waves were discussed, we see that the divergence equations play an important part in the determination of the scalar potentials.

\section{The metric and its causal structure}\label{latres}

A sufficiently simple model of a spacetime having a strong curvature singularity can be constructed by considering in $\mathbb{R}^4$ metrics of the form
\begin{equation}\label{mettoy-1}
    ds^2=-x^{2p}dt^2+dx^2+dy^2+dz^2\,,
\end{equation}
where $p$ is a real number. This metric produces only two nonzero components of the Ricci tensor, namely
\begin{equation}
    R_{tt}=(-1+p)\,p\, x^{2(-1+p)},\,\,\,\, R_{xx}=(1-p)\,p \,x^{-2},
\end{equation}
and a scalar curvature given by
\begin{equation}\label{elrtoy}
    R=g^{\mu\nu}R_{\mu\nu}=2(1-p)\, p \,x^{-2}.
\end{equation}
Additionally, the quadratic scalars $\mathcal{R}^2=R^{\mu\nu}R_{\mu\nu}$ and $\mathcal{K}=R_{\mu\nu\rho\sigma}R^{\mu\nu\rho\sigma}$ read
\begin{equation}\label{escatoy}
   \mathcal{R}^2=\mathcal{K}=2R^2.
\end{equation}
Apart from the two choices $p=0$ and $p=1$ leading to flat spacetime (in the latter, the coordinate change $x^0=x\,\sinh{(t)}$, $x^1=x\,\cosh{(t)}$, converts the $t-x$ sector of the metric given by Eq. (\ref{mettoy-1}) into the form $ ds^2=-(dx^0)^2+(dx^1)^2$), the invariants in Eqs. (\ref{elrtoy}) and (\ref{escatoy}) reveal the presence of a curvature singularity at $x=0$, which is naked.

\begin{figure}
    \centering
    \includegraphics[width=.9\linewidth]{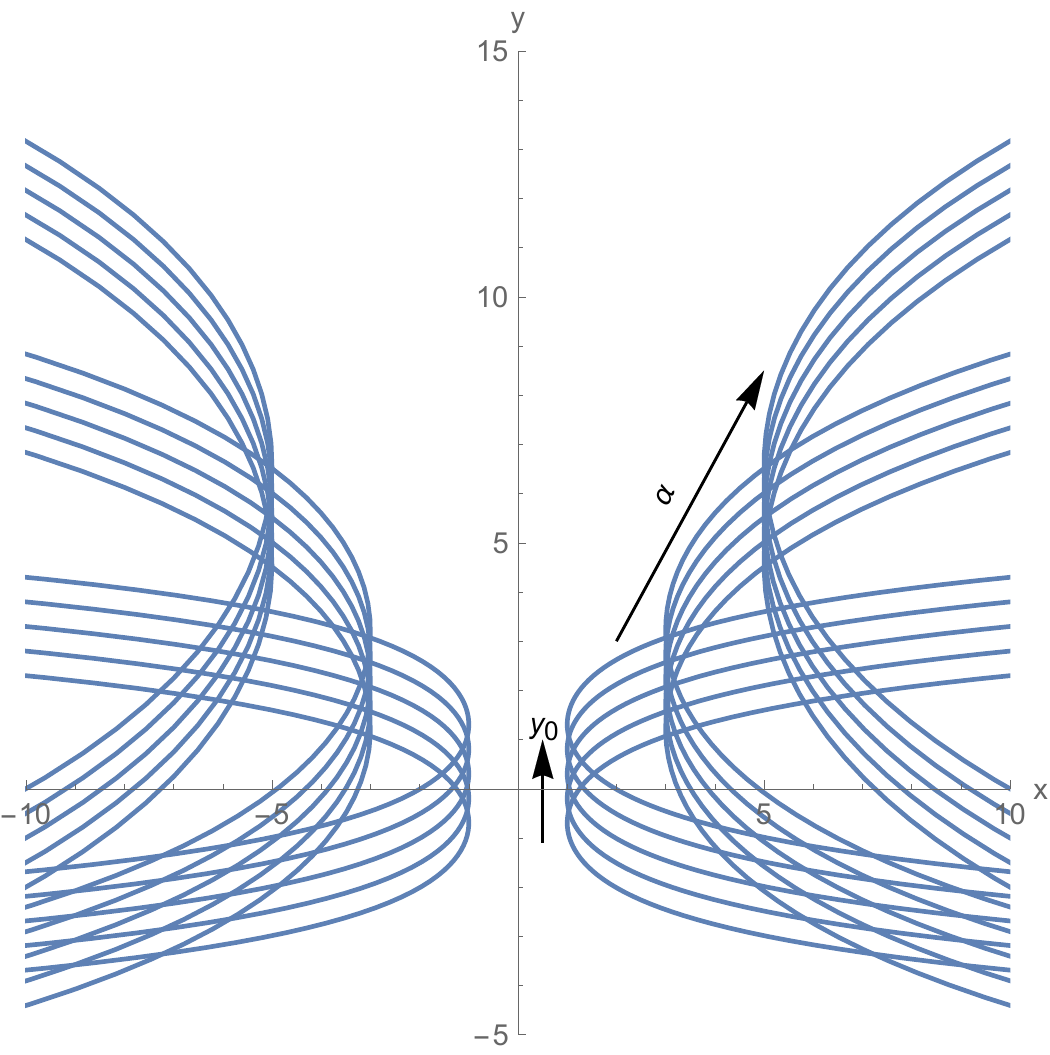}
    \caption{Null geodesic trajectories in the $(x,y)$ plane, for various values of the involved parameters, see the text.}\label{fig1}
\end{figure}

Just to make things precise, from now on we will fix $p=-1$; in this case the spacetime has negative scalar curvature, see Eq. (\ref{elrtoy}). In what follows we will study the causal geodesics and, thus, the causal structure of the space. The only two non\correction{vanishing} Christoffel symbols are
\begin{equation}\label{christoy}
    \Gamma^{t}_{\,\,\,tx}=-x^{-1},\,\,\,\,\Gamma^{x}_{\,\,\,tt}=-x^{-3}, 
\end{equation}
which lead to the geodesic equations
\begin{align}
 &\frac{d^2y}{d\,\tau^2}=\frac{d^2z}{d\,\tau^2}=0\,,\label{eq:geod_y}\\
    &\frac{d^2t}{d\,\tau^2}-\frac{2}{x}\frac{d\,x}{d\,\tau}\frac{d\,t}{d\,\tau}=0\,,\,\,\,\,\,
    \frac{d^2x}{d\,\tau^2}-\frac{1}{x^3}\left(\frac{d\,t}{d\,\tau}\right)^2=0\,,\label{eq:geod_x}
\end{align}
where $\tau$ is the affine parameter. From the first two we have  
\begin{equation}
    y(\tau)=y_0+\alpha\tau\,,\,\,\,\, z(\tau)=z_0+\beta\tau\,, \label{eq:y_tau_geodesicas}
\end{equation}
where $y_0$, $z_0$, $\alpha$, and $\beta$ are arbitrary constants. In the null case, $ds^2=0$ translates into the condition
\begin{equation}
    \frac{1}{x^2}\left(\frac{d\,t}{d\,\tau}\right)^2=\alpha^2+\beta^2+\left(\frac{d\,x}{d\,\tau}\right)^2\,.
    \label{eq:geod_nulas}
\end{equation}
Combining Eqs. (\ref{eq:geod_x}) and (\ref{eq:geod_nulas}) we get 
\begin{equation}
    x(\tau)\frac{d^2x}{d\,\tau^2}-\left(\frac{d\,x}{d\,\tau}\right)^2=\alpha^2+\beta^2\,,
\end{equation}
which results in
\begin{equation}
    x(\tau)=\pm\left(\operatorname{e}^{c_1( \tau+c_{2})}+\frac{\alpha^2+\beta^2}{4\,c_1^2}\operatorname{e}^{-c_1(\tau+c_{2})}\right)\,,
    \label{eq:x_tau_geodesicas}
\end{equation}
where $c_1$ is a constant related to the energy of the photon, \correction{and $c_{2}$ represents an irrelevant offset in $\tau$, \correction{which} why it will be omitted hereafter}. If $\alpha^2+\beta^2\neq 0$, $x(\tau)$ has a nonzero minimum value at a finite $\tau$. In turn, if $\alpha^2+\beta^2=0$, the function in Eq. (\ref{eq:x_tau_geodesicas}) becomes an exponential, which implies that $x=0$ is not attainable at any finite $\tau$. This fact indicates that the spacetime is null geodesically complete, because $x=0$ is only hit asymptotically in $\tau$. In Fig. \ref{fig1} different geodesic groups are shown as trajectories in the $(x,y)$ plane (this corresponds to $\beta=0$ in Eqs. (\ref{eq:y_tau_geodesicas}) and (\ref{eq:x_tau_geodesicas})), for the choice $c_1=1$. Each group is characterized by different choices of $y_{0}$ in Eq. (\ref{eq:y_tau_geodesicas}), while the connection between different groups is made by varying $\alpha$ as it appears in Eq. (\ref{eq:x_tau_geodesicas}).

Let us take note, for instance, that the nongeodesic, null curves $x^{\mu}(\tau)$ defined by constant values of $y$ and $z$, having
\begin{equation}\label{curnula}
t(\tau)=t_{1}^2\,\tau^2/2+t_{0},\,\,\,\, x(\tau)=t_{1}\tau,
\end{equation}
with constants $t_{0}$ and $t_{1}$, seem to reach $x=0$  \correction{(when $\tau\in[0,\tau_{0})$, the generalized affine parameter, takes $\tau=0$), and then, the curvature singularity is hit by these nongeodesic light rays. Actually, the curves described in Eq.} (\ref{curnula}) \correction{have finite generalized affine parameter length as defined by}
\begin{equation}\label{gapl}
\ell=\int_{0}^{\tau_{0}} \left[\sum_{a=0}^{3}\left(g_{\mu\nu}\frac{d x^{\mu}}{d\tau} E^{\nu}_{a}\right)^2 \right]^{1/2} d\tau\,,
\end{equation}
 \correction{where $E^{\nu}_{a}$ are the local components of a frame $E_{a}$ which is parallelly propagated along the curves $x^{\mu}(\tau)$.} This means that not every null curve is complete, only null geodesics are. 

We now proceed to characterize the timelike geodesics. As an alternative, perhaps more physical way, we can do this by using the conserved quantities associated \correction{with} the cyclic coordinates $y$, $z$. In fact, along timelike geodesics we have 
\begin{equation} \label{conser}
   p^{\nu}p_{\nu}= g_{\mu\nu}p^{\mu}p^{\nu}=-m^2,
\end{equation}
where $m$ is the mass and $p^{\mu}=m\, d x^{\mu}/d\tau$ is the four-momentum, \correction{and now $\tau$ is a parameter affinely related to proper time} . \correction{Because of} the fact that the metric does not depend on $y$ and $z$, the covariant components $p_{y}$ and $p_{z}$ are constant of motion,
\begin{align}\label{momentos}
  & p_{y}=g_{yy}\,p^{y}=g_{yy}\, m\, d y/d\tau=m \,d y/d\tau, \\
   & p_{z}=g_{zz}\,p^{z}=g_{zz}\, m\, d z/d\tau=m\, d z/d\tau.
\end{align}
Integrating, we have (compare to Eq. (\ref{eq:y_tau_geodesicas})),
\begin{equation}
    y(\tau)=y_0+\frac{p_{y}}{m}\,\tau\,,\,\,\,\,
    z(\tau)=z_0+\frac{p_{z}}{m}\,\tau\,,\label{eq:z_tau_geodesicascons}
\end{equation}
which allows us to identify $\alpha=p_{y}/m$ and $\beta=p_{z}/m$. Furthermore, we have that the energy is
\begin{equation}
   E=p_{t}=g_{tt}\,p^{t}=g_{tt}\, m \,\frac{dt}{d\tau} =-\frac{m}{x^2}\,\frac{dt}{d\tau}.
\end{equation}
From Eq. (\ref{conser}), we get
\begin{equation}
\left(\frac{dx}{d\tau}\right)^2=-1-(p_{y}/m)^2-(p_{z}/m)^2+\textbf{\textit{e}}^2\, x^2, 
\label{eq:derivada_primera_geodesicas}
\end{equation}
where $\textbf{\textit{e}}=E/m$ is the energy per unit mass. Differentiation of Eq. (\ref{eq:derivada_primera_geodesicas}) with respect to $\tau$ lead\correction{s} us to
\begin{equation}
    \frac{d^2x}{d\tau^2}=x\,\textbf{\textit{e}}^2\,,
    \label{eq:derivada_segunda_geodesicas}
\end{equation}
and directly to
\begin{equation}
    x(\tau)=d_1\operatorname{e}^{\textbf{\textit{e}}\,\tau}+d_2\operatorname{e}^{-\textbf{\textit{e}}\,\tau}\,,
    \label{eq:sol_x_geodesicas_con_ctes}
\end{equation}
where $d_1$ and $d_2$ are constants. Notice, however, that (let us say) $d_2$ is not independent; replacing Eq. (\ref{eq:sol_x_geodesicas_con_ctes}) in\correction{to} Eq. (\ref{eq:derivada_primera_geodesicas}) we obtain
\begin{equation}
    d_2=\frac{1+(p_{y}/m)^2+(p_{z}/m)^2}{4\,\textbf{\textit{e}}^2\,d_1}\,.
\end{equation}  
Finally, we can take $d_1=\pm\operatorname{e}^d$, which only represent and offset in $\tau$, and can thus be ignored. At the end we obtain
\begin{equation}
x(\tau)=\pm\left(\operatorname{e}^{\textbf{\textit{e}}\,\tau}+\frac{1+(p_{y}/m)^2+(p_{z}/m)^2}{4\,\textbf{\textit{e}}^2}\operatorname{e}^{-\textbf{\textit{e}}\,\tau}\right)\,,\label{eq:x_tau_geodesicas_temporales_2}
\end{equation}
which has a similar structure to Eq. (\ref{eq:x_tau_geodesicas}). Because $1+(p_{y}/m)^2+(p_{z}/m)^2$ is always different from zero, the geodesics reach a minimum value and then bounce back, just as the null geodesics with $\alpha^2+\beta^2\neq 0$. We proved, then, that the singularity is repulsive in what concerns the behavior of null and timelike geodesics, \correction{which is} a consequence related to the negativeness of the scalar curvature; this is very \correction{similar} to what happens in the Reissner-Nordstr\"{o}m metric, where the singularity is repulsive for timelike geodesics \cite{Maluf}, and in certain regular black hole solutions present in alternative gravity theories, where the role of the Schwarzschild singularity is played by a regular core \correction{that} repels the causal geodesics; consult, e.g., \cite{nos2} and \cite{nos1}. We see that the metric in question is not only simple enough, but also useful as a toy model for studying more physical examples where strong curvature, repulsive singularities actually appear.  

\bigskip

The global structure of the spacetime can be studied by constructing the Penrose diagram. In the $(t,x)$ plane, let us first consider the coordinate change $\chi=\pm\, x^2/2$, which puts the metric into the form  
\begin{align}
    ds^2=\frac{1}{2\chi}\left(-dt^2+d\chi^2\right)\,.
\end{align}
Then, null coordinates can be defined as $ u=\chi+t$ and $v=\chi-t$, in such a way that the interval simply looks like $ ds^2=(u+v)^{-1}du\,dv$. The compactification takes place by means of new coordinates $U$ and $V$, both ranging in the interval $[-\frac{\pi}{2},\frac{\pi}{2}]$ and defined by 
\begin{equation}
    U=\arctan(u)\,,\,\,\,\, V=\arctan(v)\,,
\end{equation}
which convert the metric into
\begin{equation}
    ds^2=\cos^{-2}(U)\cos^{-2}(V)(\tan U+\tan V)^{-1} dU\,dV\,.
    \end{equation}
One last coordinate change $X,T$ will bring the interval back again to a form conformal to Minkowski space, namely, 
\begin{align}
    X=U+V\,&,\,\,\,\, T=U-V\,,\\
    -\pi\leq X+T\leq \pi\,&,\,\,\,\, 
    -\pi\leq X-T\leq \pi\,,
\end{align}
so as to put the metric in its final form
\begin{equation}
    ds^2=\frac{1}{\omega^2}\left(-dT^2+dX^2\right)\,,
\end{equation}
with $\omega^2=\left(\cos T+\cos X\right)^2\left(\tan\left(\frac{X+T}{2}\right)+\tan\left(\frac{X-T}{2}\right)\right)$, which \correction{vanishes} at $X=0$.

\begin{figure}
    \centering
   \includegraphics[width=.9\linewidth]{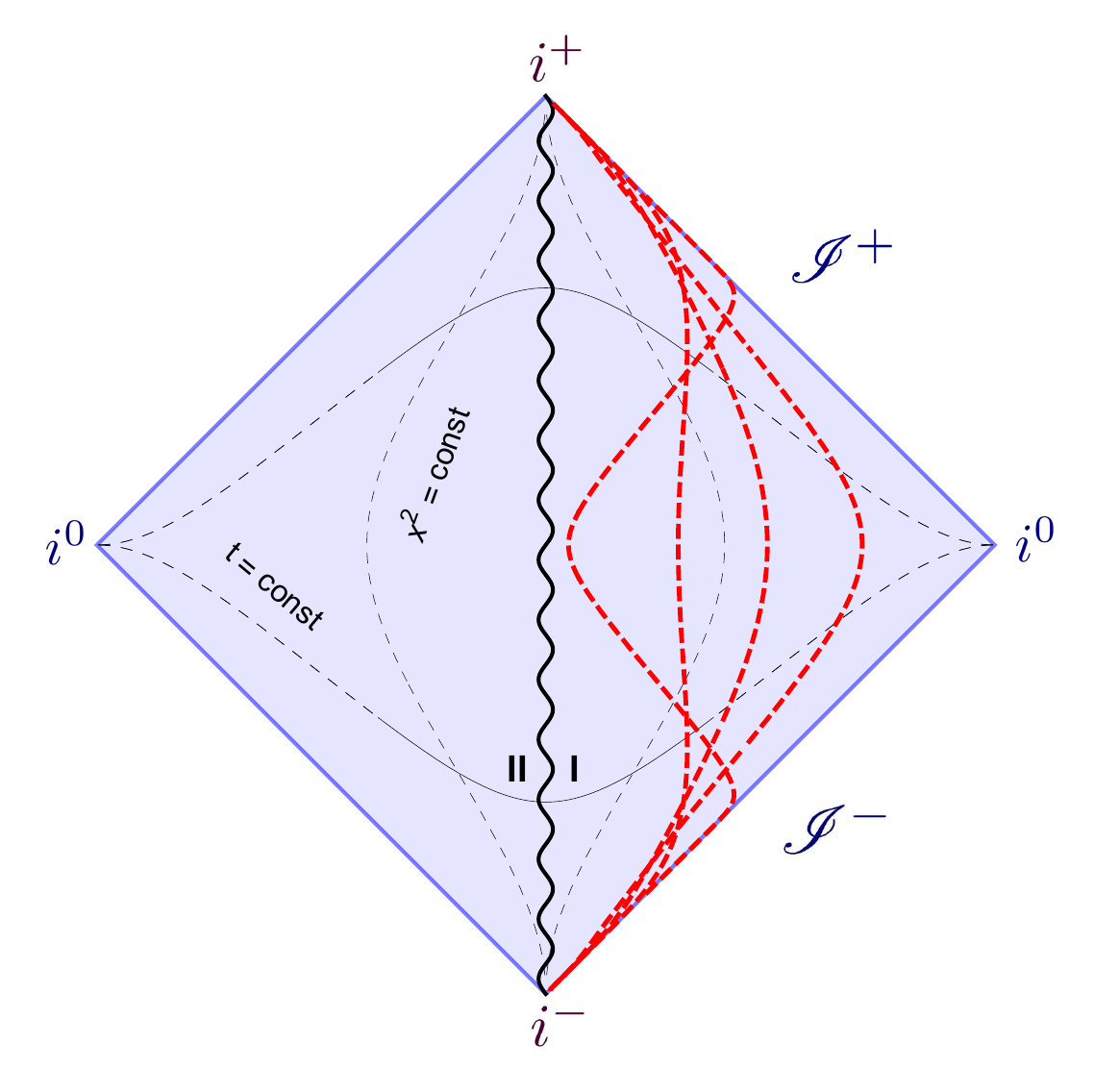}
    \caption{Penrose diagram for the metric (\ref{mettoy-1}) with $p=-1$.}
    \label{fig:diagrama-penrose-toy-model}
\end{figure}

The Penrose diagram is presented in Fig. \ref{fig:diagrama-penrose-toy-model}, where the two regions I ($x^2=2\chi$) and II ($x^2=-2\chi$) are shown. The wavy timelike curve at the center represents the naked curvature singularity. Timelike geodesics, as come from taking different values of $p_{y}/m$, $p_{z}/m$ and $\textbf{\textit{e}}$ in Eq. (\ref{eq:x_tau_geodesicas_temporales_2}), are visualized in dotted red lines. We see that the singularity is repulsive in terms of the behavior of null and timelike geodesics. However, null curves, as the one discussed in Eq. (\ref{curnula}) (not shown in the diagram), can actually reach the singularity in a finite affine parameter $\tau$. The Penrose diagram is basically the same as the one corresponding to the Reissner-Nordstr\"{o}m space in which the electric charge $Q$ is greater than the \correction{Arnowitt-Deser-Misner} mass $M$. \correction{Nonetheless, it might be worthwhile to point out that, in the latter, radial null geodesics actually reach the singularity at a finite affine parameter.}   

\section{Electrostatics}\label{estatica}

Beyond the behavior of light rays in the medium or, equivalently, of the null geodesics in the spacetime, it is interesting to study the main features regarding the structure of a more general electromagnetic field in this model. In the rest of the paper, we will focus on the static case, in particular, when no magnetic field is present. The static PT equations given by Eq. (\ref{estaticac}) for the electrostatic potential are  
\begin{equation}
    \bar{\nabla} \cdot (\textbf{K}\bar{\nabla}\phi)=0,\,\,\,\,\,\,\,\,
     (\bar{\nabla} \times\bar{\Gamma})\cdot\bar{\nabla}\phi=0,\label{estaticace}
\end{equation}
where we have written $\phi=\phi_{E}$. Using the definitions in Eq. (\ref{matyvec}) in the particular case under consideration, i.e., for the metric in Eq. (\ref{mettoy-1}) with $p=-1$, we obtain right away 
\begin{equation}\label{eq:Kcom}
\textbf{K}=|x|\,\textbf{I}\,,\,\,\,\,\,\,\bar{\Gamma} =\bar{0}\,.
\end{equation}
In this way, Eqs. (\ref{estaticace}) end up being
\begin{align}
   &\nabla^2\phi=- \frac{1}{x}\frac{\partial \phi}{\partial x}\,.\label{eq:dif_potencial}
\end{align}
Let us consider thus $\phi=\mathcal{X}(x)\mathcal{Y}(y)\mathcal{Z}(z)$, which converts Eq. (\ref{eq:dif_potencial}) into  
\begin{align}
&x\,\mathcal{X}''+\mathcal{X}'+a^2\,x\,\mathcal{X}=0\,,\\
    &\mathcal{Y}''+b^2 \mathcal{Y}=0\,,\\
    &\mathcal{Z}''-c^2\mathcal{Z}=0\,,
\end{align}
where $a$ and $b$ are separation constants, and $c^2=a^2+b^2$. The set of solutions is grouped according to
\begin{equation} \label{chidex}
    \mathcal{X}_{a}(x)=\left\{\begin{array}{lcc}
        A_0+\tilde{A}_0\,\ln(x) & \mbox{if } & a=0\,,\\
        A_a \,J_0(a\, x)+ \tilde{A}_a \,Y_0(a\, x) & \mbox{if } & a\neq0\,,
    \end{array}\right.
\end{equation}
\begin{equation} \label{partey}
    \mathcal{Y}_b(y)=\left\{\begin{array}{lcc}
        B_0+\tilde{B}_0\,y & \mbox{if } & b=0\,, \\
        B_b\,\operatorname{e}^{i\,b\,y}+\tilde{B}_b\,\operatorname{e}^{-i\,b\,y} & \mbox{if } & b\neq0\,,
    \end{array}\right.
\end{equation}
\begin{equation}\label{partez}
    \mathcal{Z}_{c}(z)=\left\{\begin{array}{lcc}
        C_0+\tilde{C}_0\,z & \mbox{if } & c=0\,, \\
        C_c\,\operatorname{e}^{c\,z}+\tilde{C}_c\,\operatorname{e}^{-c\,z} & \mbox{if } & c\neq0\,,
    \end{array}\right.
\end{equation}
where $J_{0}(ax)$ and $Y_{0}(ax)$ are Bessel functions of first and second kind, respectively. Moreover, all the capitals $A$, $\tilde{A}$, etc. are constants. In this manner, the potential in Eq. (\ref{eq:dif_potencial}) is really a combination of modes
\begin{equation}\label{eq:phi_general}
    \phi(x,y,z)=\sum_{a,b}  \mathcal{X}_a(x) \mathcal{Y}_b (y)\mathcal{Z}_{c}(z)\,.
\end{equation}
It is clear that divergences at $x=0$ generally occur in $\mathcal{X}_a(x)$. The situation is not different from having divergences at $r=0$ when considering waves in a usual, hollow cylindrical waveguide, and they are ultimately a consequence of the structure of the Laplace operator in cylindrical coordinates. However, in the cylindrical waveguide the divergent modes are willfully dismissed on physical grounds, for they lead to an experimentally unobserved divergent power flux in the guiding system. Nonetheless, in our present circumstances, divergent modes should not be discarded from scratch. Actually, there are no physical reasons to exclude divergent solutions when the very structure of the spacetime is singular. On the contrary, what seems more striking is the possibility of having a regular electric field on top of the curvature singularity. In what follows we will delve into this fact. 

Note that the divergence at $x=0$ of the mode expansion in Eq. (\ref{eq:phi_general}) is due to the presence of the logarithmic term and the function $Y_0(a x)$ in Eq. (\ref{chidex}). Therefore, the regularity of the potential will be ensured if both $\tilde{A}_{0}$ and all the $\tilde{A}_a$ are zero. Additionally, we can consider the identification $x=x_{0}$ with $x=-x_{0}$, which corresponds to folding the material in the $x$ direction. This kind of topological control, usually conceived only on formal grounds in the spacetime ``side'' of the analogy, is perfectly feasible in the \correction{laboratory}. This implies that $\mathcal{X}_a(x_{0})=\mathcal{X}_a(-x_{0})$ and $\mathcal{X}'_a(x_{0})=\mathcal{X}'_a(- x_{0})$. The first condition is satisfied for any $x_{0}$ since $J_0$ is an even function. As a result, its derivative is odd, and thus the boundary condition on the derivative implies $\mathcal{X}'_a(x_{0})=\mathcal{X}'_a(-x_{0})=0$. Hence, $a^2\geq0$. In particular, since $J_{0}'(a x)=-J_{1}(a x)$, we have $a\equiv a_n=j_{1,n}/x_{0}$, where $j_{1,n}$ is the $n$th zero of $J_1$. Since $J_0(ax)$ is an even function, it is sufficient to consider $n>0$.

To better visualize the structure of the potential, we could first consider a \correction{two}-dimensional  \correction{analog} medium by taking constant $z$ surfaces. This is done without loss of generality due to the translational symmetry in the $y$ and $z$ directions. In this case, we will have only one separation constant, since $b^2=-a^2\leq0$. Thus, by identifying $x=x_{0}$ with $x=-x_{0}$, the $(x,y)$ plane folds to form a cylindrical surface where the potential results 
\begin{align}
\phi(x,y)=B_0\,y+\sum_{a} A_a\,J_0(ax)\left(\operatorname{e}^{-a\,y}+B_a\,\operatorname{e}^{a\,y}\right)\,,
\label{eq:potencial_electrico_2_dimensiones}
\end{align}
where, remember, $a\equiv a_{n}=j_{1,n}/x_{0}$. In the expression given by Eq. (\ref{eq:potencial_electrico_2_dimensiones}), an irrelevant additive constant was omitted, and others were conveniently redefined. The omission of an additive constant in Eq. $(\ref{eq:potencial_electrico_2_dimensiones})$ simply fixes the potential at the origin according to $ \phi(0,0)=\sum_{a} A_a(1+B_{a})$. 

In order to fully determine the potential (i.e., the constants $A_a$, $B_0$ and $B_a$), we need to specify boundary conditions on a closed curve limiting the two-dimensional surface. Given that we have identified $x=x_{0}$ with $x=-x_{0}$, we can specify the potential $\phi(x,y)$, for instance, along two lines of constant $y$, by setting $\phi(x,y_1)=\phi_1(x)$ and $\phi(x,y_2)=\phi_2(x)$. Therefore, using the orthonormality relations
\begin{align}
    &\int_0^{x_{0}} x\,J_0\left(a_{n}x\right)J_0\left(a_{m}x\right)dx 
    =\delta_{nm}\frac{x_{0}^2}{2}J_0^2(a x_{0})\equiv\delta_{nm}\tilde{x}_{0}^2  ,\notag\\
   & \int_0^{x_{0}} x\,J_0\left(a x\right)\,dx=0\,,\label{eq:ortogonalidad_bessel_2}
\end{align}
where we defined $\tilde{x}_{0}^2=x_{0}^2J_0^2(a x_{0})/2$, we have

\begin{align}
    \int_0^{x_{0}} x&J_0\left(ax\right)\phi_1(x)dx=A_a\tilde{x}_{0}^2
\left(\operatorname{e}^{-a\,y_1}+B_a\,\operatorname{e}^{a\,y_1}\right),\label{const100}\\
    \int_0^{x_{0}} x\,&\phi_1(x)\,dx=B_{0}\,y_1 x_{0}^2/2,\label{const10}\\
    \int_0^{x_{0}} x&J_0\left(ax\right)\phi_2(x)dx=A_{a}\tilde{x}_{0}^2
    \left(\operatorname{e}^{-a\,y_2}+B_a\,\operatorname{e}^{a\,y_2}\right),\label{const200}\\
    \int_0^{x_{0}} x\,&\phi_2(x)\,dx=B_0\,y_2x_{0}^2/2\,.\label{eq:ctes_campo_elctrostatico_2D_d}
\end{align}
By using these relations, we can solve for all the constants simply by knowing either $\phi_1(x)$ or $\phi_2(x)$. Precisely, from Eqs. (\ref{const10}) and (\ref{eq:ctes_campo_elctrostatico_2D_d}) we see that 
\begin{equation}\label{vinculo}
    \frac{1}{y_{2}}\int_0^{x_{0}} x\,\phi_2(x)\,dx= \frac{1}{y_{1}}\int_0^{x_{0}} x\,\phi_1(x)\,dx\,,
\end{equation}
so there is a link between $\phi_1(x)$ and $\phi_2(x)$, which ultimately comes from the fact that the potential was fixed at the origin; see the comments on the paragraph below Eq. (\ref{eq:potencial_electrico_2_dimensiones}). In particular, if (let us say) $\phi_1(x)$ is constant, we have the relation $B_{0}=\phi_1/y_{1}=\phi_2/y_{2}$, which fixes the constant value of $\phi_2$. Additionally, from Eqs. (\ref{const100}) and (\ref{const200}) we see that $A_{a}=0$ for all $a$ in this case. We end up, thus, with a linear potential and a constant electric field of the form $\bar{E}_{0}=-B_{0}\hat{y} $. This represents the first example of a regular electrostatic field, even in the presence of a curvature singularity.

Let us elaborate \correction{further with} another, more involved example. For instance, consider $x_{0}=3$, $y_1=-1$, $y_2=1$, and $\phi_1(x)=x-2x_{0}/3$. This functional form for $\phi_1(x)$, according to Eq. (\ref{const10}), conduces to a \correction{vanishing} $B_{0}$ and fixes the form of $\phi_2(x)$ to be just $\phi_2(x)=\pm\phi_1(x)$, as Eq. (\ref{vinculo}) demands. The constants $A_{a}$ and $B_{a}$ are determined by means of the integrals of Eqs. (\ref{const100}) and (\ref{const200}), where we have taken $\phi_2(x)=\phi_1(x)$. At the end all the constants read
\begin{align}
&B_{0}=0\,,\label{consf1}\\
&A_a=A_{a_{n}}=\frac{_{\textbf{1}}F_{\textbf{2}}\left(\{\frac{3}{2}\},\{1,\frac{5}{2}\},-\frac{j_{1,n}^2}{4}\right)}{\cosh{(j_{1,n}/3)}\,J_0^2(j_{1,n})}\hspace{0.5cm}(n\geq1)\,,\label{consf2}\\
    &B_a=B_{a_{n}}=1\hspace{0.5cm}(n\geq1)\,,\label{consf3}
\end{align}
where $_{\textbf{p}}F_{\textbf{q}}$ is the generalized hypergeometric function \cite{hiper}, given by
\begin{equation}
    _{\textbf{p}}F_{\textbf{q}}(\{b_1,...,b_p\},\{c_1,...,c_q\},z)=
    \sum_{s=0}^\infty\frac{(b_1)_s...(b_p)_s}{(c_1)_s...(c_q)_s}\frac{z^s}{s!},\notag
\end{equation}
which is defined in terms of the rising factorial
\begin{align}
    &(b)_0=1\,,\notag\\
    &(b)_s=b(b+1)(b+2)\dots(b+s-1)\hspace{0.5cm}(s\geq1)\,.\notag
\end{align}
\begin{figure}
    \centering
    \begin{subfigure}{.75\linewidth}
    \centering
        \includegraphics[width=\linewidth]{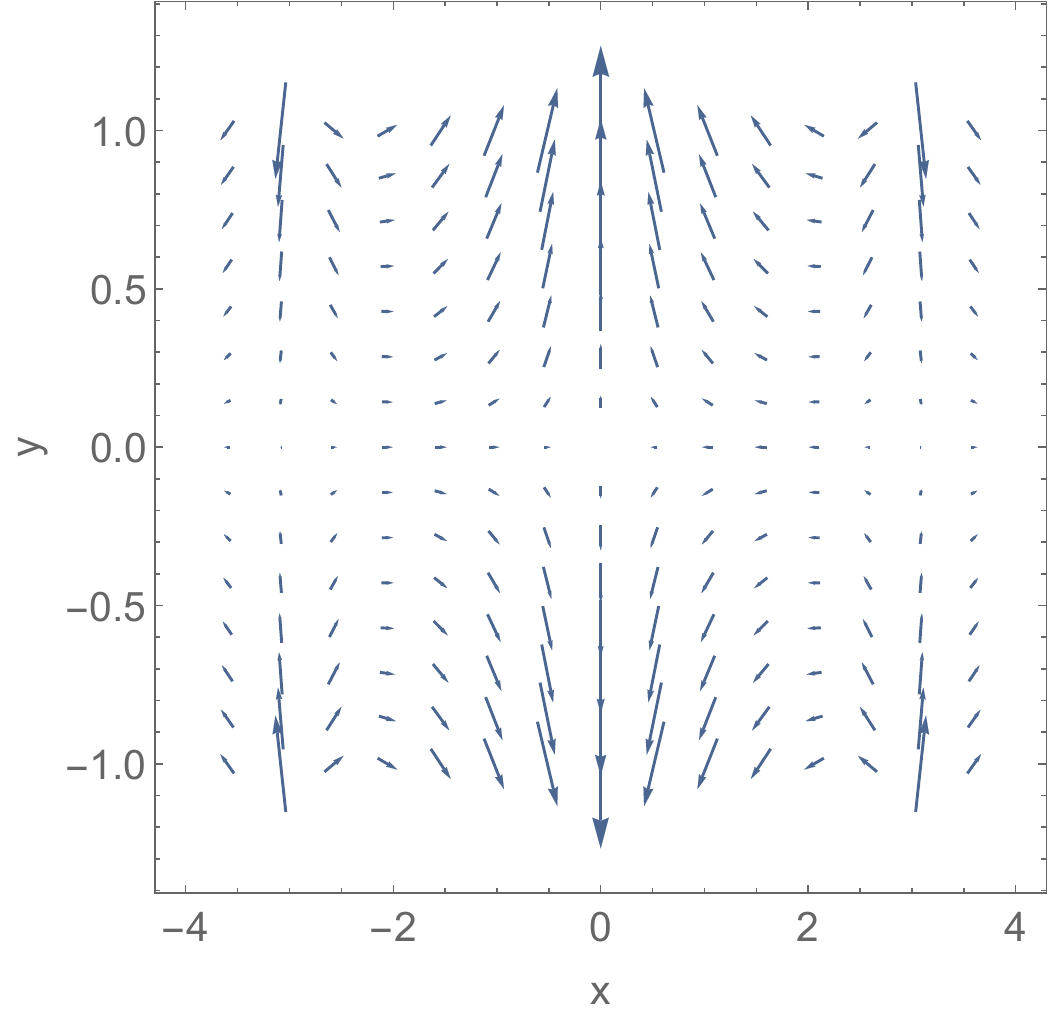}
        \caption{\label{subfig:electrostatica_plano}}
    \end{subfigure} \\
  
    \begin{subfigure}{.8\linewidth}
    \centering
        \includegraphics[width=\linewidth]{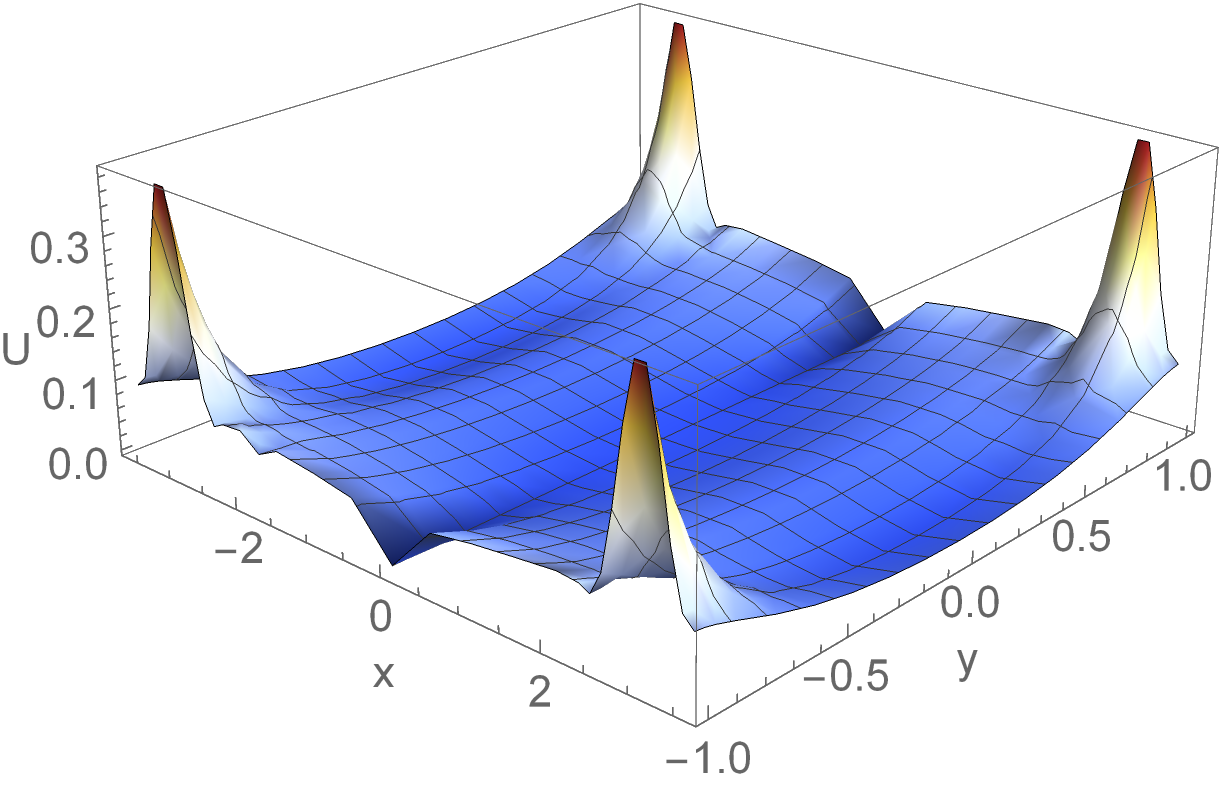}
        \caption{\label{subfig:energia_electrostatica}}
    \end{subfigure}
    \caption{(\subref{subfig:electrostatica_plano}) Vector plot of the electrostatic field corresponding to the potential given by Eq. (\ref{eq:potencial_electrico_2_dimensiones}), with the constants given in Eqs. (\ref{consf1})-(\ref{consf3}). (\subref{subfig:energia_electrostatica}) Electrostatic energy density $U(x,y)$ for the same case.}
    \label{fig:campo-electrostatica}
\end{figure}
In Fig. \ref{fig:campo-electrostatica}(\subref{subfig:electrostatica_plano}) we show a vector plot of the electrostatic field $\bar{E}=-\bar{\nabla} \phi$ in the case just considered. Not only \correction{is} the electric field regular (actually, \correction{zero}) at the origin, but along the entire $y$ axis; its functional form there reads
\begin{equation}
\lim_{{x\to 0}}\bar{E}(x,y)\propto\sum_{n=1}^{\infty} j_{1,n}\,A_{a_{n}}\,\sinh\left({\frac{j_{1,n}}{x_{0}}\,y}\right)\,\hat{y}\,. 
\end{equation}
In Fig. \ref{fig:campo-electrostatica}(\subref{subfig:energia_electrostatica}) we show the electrostatic energy density $U(x,y)\propto\bar{D}\cdot\bar{E}=\textbf{K}\bar{E}\cdot\bar{E}= |x||\bar{E}|^2$. As a consequence of the regularity of the electric field, the energy density is \correction{zero} along the entire line $x=0$. We conclude that the usual devastating effects associated \correction{with} the presence of curvature singularities are not such in this example.  

\bigskip

After the preceding two-dimensional digression, we can go back to the three-dimensional case and sketch the main steps leading to the full determination of the electrostatic potential. To start, we can now introduce an additional (and optional) topological condition by identifying $y=y_{0}$ with $y=-y_{0}$. This forces $b^2\geq0$, $\tilde{B}_0=0$, and $B_b=\tilde{B}_b$ in Eq. (\ref{partey}), which conduces to $\mathcal{Y}(y)\propto \cos\left(b \,y\right)$, where $b\equiv b_{m}=m\pi/y_{0}$, and $m$ is a positive integer. Therefore, the expansion given by Eq. (\ref{eq:phi_general}) reads
\begin{equation}
\phi=B_{00}z+\sum_{n,m=1}^\infty A_{nm}J_0\left(ax\right)\cos\left(b\,y\right)
    \left(\operatorname{e}^{-c z}+B_{nm}\,\operatorname{e}^{cz}\right)\,,\nonumber
\end{equation}
where $c\equiv c_{nm}=\sqrt{a_{n}^2+b_{m}^2}$. Just as in the two-dimensional case worked before, the constants $B_{00}$, $A_{nm}$ and $B_{nm}$ can be determined by specifying the potential on two sections of constant $z$, this is, $\phi(x,y,z_1)=\phi_1(x,y)$ and $\phi(x,y,z_2)=\phi_2(x,y)$. Hence, using 3D orthonormality conditions similar to those in Eq. (\ref{eq:ortogonalidad_bessel_2}), we get 
\begin{align}
\int_0^{x_{0}} \int_0^{y_{0}}x& J_0\left(a x\right)\cos\left(b\,y\right)\phi_1(x,y)dx \,dy\notag \\
&=A_{nm}\,y_{0}\,\tilde{x}_{0}^2\left(\operatorname{e}^{-c\, z_1}+B_{nm}\,\operatorname{e}^{c\, z_1}\right)/2\,,
\end{align}
\begin{align}
    &\int_0^{x_{0}}\int_0^{y_{0}} \,x\,\phi_1(x,y) dx\, dy=B_{00}\,z_1\,x_{0}^2\,y_{0}/2\,,
\end{align}
\begin{align}
\int_0^{x_{0}} \int_0^{y_{0}}x& J_0\left(a x\right)\cos\left(b\,y\right)\phi_2(x,y)dx \,dy\notag \\
&=A_{nm}\,y_{0}\,\tilde{x}_{0}^2\left(\operatorname{e}^{-c\, z_2}+B_{nm}\,\operatorname{e}^{c\, z_2}\right)/2\,,
\end{align}
  \begin{align}  
     &\int_0^{x_{0}}\int_0^{y_{0}} \,x\,\phi_2(x,y) dx\, dy=B_{00}\,z_2\,x_{0}^2\,y_{0}/2\,,
\end{align}
 from which we can solve for the unknown constants. 
 
\bigskip

\section{Concluding comments}
In this work, we have studied the behavior of the electrostatic field in a spacetime hosting a strong, naked curvature singularity using the Plebanski-Tamm formalism. We have found that there are exact solutions to the full electrostatic equations that remain bounded in any neighborhood of the curvature singularity. These solutions describe nontrivial electrostatic fields \correction{that result in a zero} electrostatic energy as $x \rightarrow 0$, \correction{which is in stark contrast to what occurs with extended bodies in that limit, due to the unbounded tidal forces generated by the curvature singularity}. Although the metric employed, \emph{per se}, may not hold the most physical relevance, in the sense that it does not describe a solution of the Einstein equations with any reasonable matter content (though it nicely models the repulsive singularity existing, for instance, in the Reissner-Nordstr\"{o}m metric), its ability to sustain a regular electric field is highly significant. More than discussing a certain solution, our aim was to delve into the more general possibility of having a regular field in a singular space of what can be considered the worst type. It is sounding increasingly likely that this sort of regularity could be sustained in a dynamical situation in which the electromagnetic energy might be able to flow from one side to the other, through the singularity. This possibility constitutes material of an ongoing investigation.

\bigskip

\textbf{\emph{Acknowledgements.}} The authors want to express their gratitude to D. Grosz and S. Hernández for examining the manuscript and contributing with keen, constructive suggestions. F. F. is member of \emph{Carrera del Investigador Cient\'{i}fico} (CONICET). This work has been supported by CONICET and Instituto Balseiro (UNCUYO).

\appendix
\section{Consistency of the system given by Eqs. (\ref{eq:rot_H_cuasi_planas}) and (\ref{eq:div_B_igual_0}) }\label{ap1}
Here we prove that the Eqs. (\ref{eq:rot_H_cuasi_planas}) alone are sufficient to determine everything that is possible to be determined within the context of the quasi-plane\correction{-wave} ansatz written in Eq. (\ref{expEyH2}). From the second Eq. (\ref{eq:rot_H_cuasi_planas}), we have
\begin{equation}
\textbf{K} \bar{E}_{0}=i\,k_{0}^{-1}\,\bar{\nabla}\times\bar{H}_{0}-(\bar{\nabla}(\bar{k}\cdot\bar{x})+\bar{\Gamma})\times\bar{H}_{0}\label{eq:rot_H_cuasi_planasap}\,.
\end{equation}
We can plug this into the first Eq. (\ref{eq:div_B_igual_0}), this is  
\begin{widetext}
\begin{align}
   -i\,k_0\left[\bar{\nabla}(\bar{k}\cdot\bar{x})+\bar{\Gamma}\right]\cdot \Big[i\,k_{0}^{-1}\,\bar{\nabla}\times\bar{H}_{0}-(\bar{\nabla}(\bar{k}\cdot\bar{x})+\bar{\Gamma})\times\bar{H}_{0} \Big] -(\bar{\nabla}\times\bar{\Gamma})\cdot\bar{H_0}
   = \bar{\nabla}\cdot\Big[i\,k_{0}^{-1}\,\bar{\nabla}\times\bar{H}_{0}-(\bar{\nabla}(\bar{k}\cdot\bar{x})+\bar{\Gamma})\times\bar{H}_{0} \Big], \label{eq:div_B_igual_0ap}
\end{align}
\end{widetext}
\correction{Because of} the fact that the vector $(\bar{\nabla}(\bar{k}\cdot\bar{x})+\bar{\Gamma})\times\bar{H}_{0}$ is orthogonal to $\bar{\nabla}(\bar{k}\cdot\bar{x})+\bar{\Gamma}$, the LHS of Eq. (\ref{eq:div_B_igual_0ap}) reads 
\begin{equation}
 \left[\bar{\nabla}(\bar{k}\cdot\bar{x})+\bar{\Gamma}\right]\cdot \left[\bar{\nabla}\times\bar{H}_{0} \right] -(\bar{\nabla}\times\bar{\Gamma})\cdot\bar{H_0}\,.\label{inter2}
\end{equation}
On the other hand, by virtue that the divergence of a curl identically vanish, the RHS of Eq. (\ref{eq:div_B_igual_0ap}) ends up being  
\begin{equation}
-\bar{\nabla}\cdot\Big[(\bar{\nabla}(\bar{k}\cdot\bar{x})+\bar{\Gamma})\times\bar{H}_{0} \Big].\label{inter}
\end{equation}
Now, we can make use of the identity
 \begin{equation}
\bar{\nabla}\cdot (\bar{A}\times\bar{B})=(\bar{\nabla}\times\bar{A})\cdot\bar{B}- \bar{A}\cdot(\bar{\nabla}\times\bar{B}),  
 \end{equation}
to convert Eq. (\ref{inter}) into
\begin{equation}
-\left[\bar{\nabla}\times(\bar{\nabla}(\bar{k}\cdot\bar{x})+\bar{\Gamma})\right]\cdot \bar{H_0}+\left[\bar{\nabla}(\bar{k}\cdot\bar{x})\times\bar{\Gamma}\right]\cdot\left[\bar{\nabla}\times\bar{H_0}\right].\label{inter3}
\end{equation}
Equating Eqs. (\ref{inter2}) and (\ref{inter3}) we get 
\begin{equation}
  (\bar{\nabla}\times\bar{\Gamma})\cdot\bar{H_0}
   = \left[\bar{\nabla}\times(\bar{\nabla}(\bar{k}\cdot\bar{x})+\bar{\Gamma})\right]\cdot \bar{H_0}, 
\end{equation}
or
 \begin{equation}  
(\bar{\nabla}\times\bar{\Gamma})\cdot\bar{H_0}=\left[\bar{\nabla}\times\bar{\nabla}(\bar{k}\cdot\bar{x})+  \bar{\nabla}\times\bar{\Gamma}\right]\cdot \bar{H_0}.\label{inter4}
\end{equation}
However, because the curl of a gradient identically vanish\correction{es},  $\left[\bar{\nabla}\times\bar{\nabla}(\bar{k}\cdot\bar{x})\right]\cdot \bar{H_0}=0$, Eq. (\ref{inter4}) produces an identity. This means that the combination of the second Eq. (\ref{eq:rot_H_cuasi_planas}) and the first Eq.  (\ref{eq:div_B_igual_0}) do not provide more information than the one encoded in Eq. (\ref{eq:rot_H_cuasi_planas}) alone. Of course, the same conclusion can be obtained by starting from the first Eq. (\ref{eq:rot_H_cuasi_planas}) and combining it with the second Eq.  (\ref{eq:div_B_igual_0}).

\end{document}